%% file: main_revision.tex
\documentclass[journal,twoside,web]{ieeecolor}
\usepackage{jsen}
\usepackage{cite}
\usepackage{amsmath,amssymb,amsfonts}
\usepackage{algorithmic}
\usepackage{graphicx}
\usepackage{textcomp}
\usepackage{wrapfig}

\usepackage{graphicx}
\usepackage{dcolumn}
\usepackage{bm}
\usepackage{multirow}
\usepackage[exponent-product = \cdot]{siunitx}
\usepackage[nolist,nohyperlinks]{acronym} 
\usepackage{xcolor}
\definecolor{ibilight}{RGB}{193,216,237}
\definecolor{ibidark}{RGB}{0,73,146}	
\definecolor{uke2}{RGB}{170,156,143} 	
\definecolor{uke3}{RGB}{87,87,86}		
\definecolor{ukesec1}{RGB}{255,223,0}	
\definecolor{ukesec2}{RGB}{239,123,5}	
\definecolor{ukesec3}{RGB}{104,195,205}	
\definecolor{ukesec4}{RGB}{138,189,36}	
\definecolor{tuhh}{RGB}{45,198,214}     

\usepackage{tikz}
\usepackage{siunitx}
\DeclareSIUnit{\sample}{S}
\usepackage{makecell}
\usepackage{array}

\def\BibTeX{{\rm B\kern-.05em{\sc i\kern-.025em b}\kern-.08em
    T\kern-.1667em\lower.7ex\hbox{E}\kern-.125emX}}
\definecolor{abstractbg}{rgb}{0.89804,0.94510,0.83137}
\PassOptionsToPackage{
    colorlinks,
    citecolor=ibidark,
    linkcolor=ibidark,
    urlcolor=ibidark
}{hyperref}
\usepackage{hyperref} 
\usepackage[all]{hypcap} 

\begin{document}
\title{Natural Frequency Dependency of \\Magneto-Mechanical Resonators on \\ Magnet Distance}
\author{Jonas Faltinath, Fabian Mohn, Fynn Foerger, Martin Möddel, and Tobias Knopp
\thanks{Date: \today}
\thanks{The authors are with the Section for Biomedical Imaging, University Medical Center Hamburg-Eppendorf, Hamburg, Germany and with the Institute for Biomedical Imaging, Hamburg University of Technology, Hamburg, Germany (e-mail: j.faltinath@uke.de)}
\thanks{T. Knopp is also with the Fraunhofer Research Institution for Individualized and Cell-Based Medical Engineering IMTE, Lübeck, Germany}
\thanks{This work was supported by the Deutsche Forschungsgemeinschaft (DFG, German Research Foundation), SFB 1615, Project number 503850735 and 555456057.}}

\IEEEtitleabstractindextext{%
\fcolorbox{abstractbg}{abstractbg}{%
\begin{minipage}{\textwidth}%
\begin{wrapfigure}[20]{r}{2.8in}%
\includegraphics[width=2.6in]{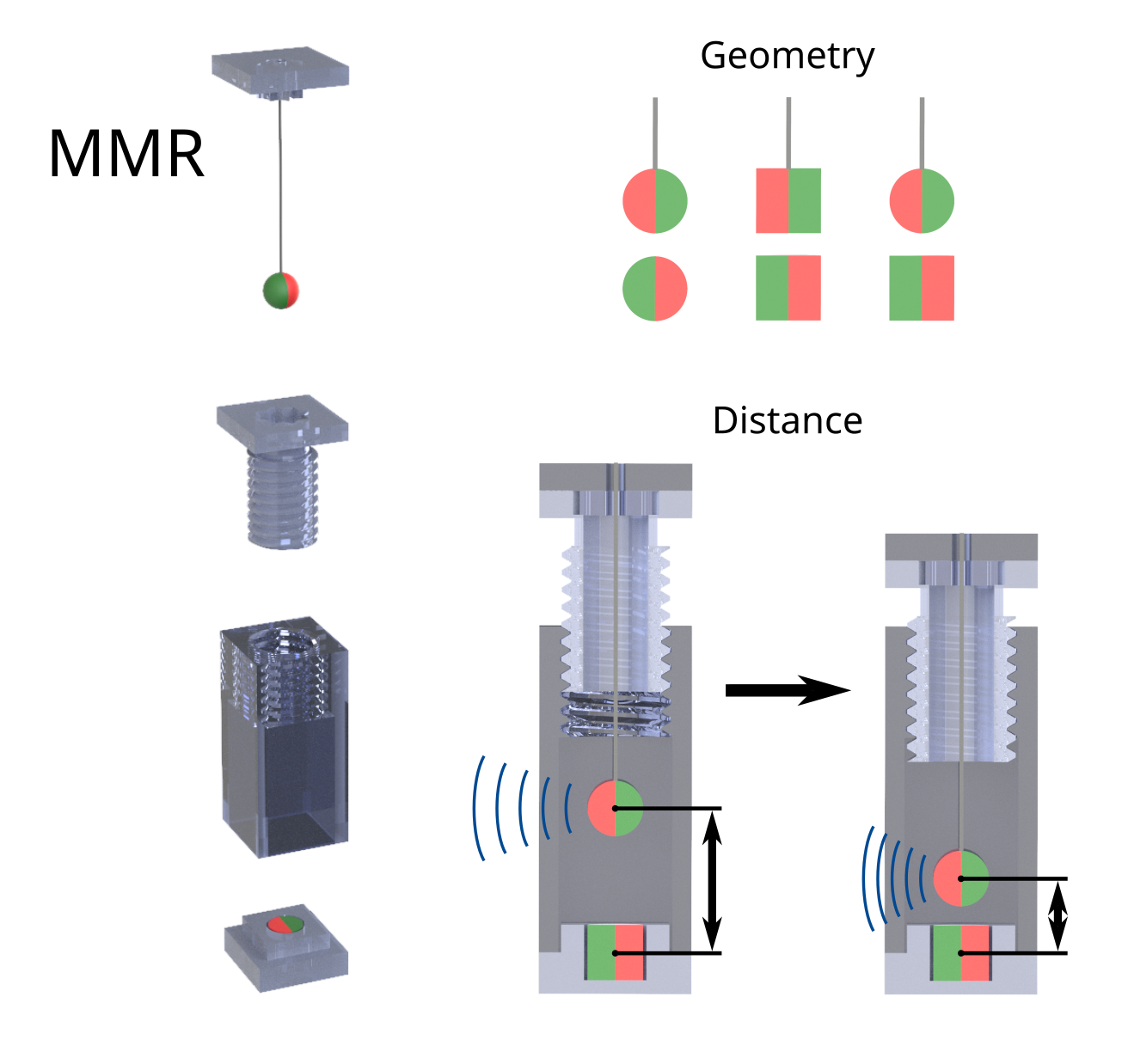}%
\end{wrapfigure}%
\begin{abstract}
The precise derivation of physical quantities like temperature or pressure at arbitrary locations is useful in numerous contexts, e.g.~medical procedures or industrial process engineering.
The novel sensor technology of magneto-mechanical resonators (MMR), based on the interaction of a rotor and stator permanent magnet, allows for the combined tracking of the sensor position and orientation while simultaneously sensing an external measurand.
Thereby, the quantity is coupled to the torsional oscillation frequency, e.g.~by varying the magnet distance. 
In this paper, we analyze the (deflection angle-independent) natural frequency dependency of MMR sensors on the rotor-stator distance, and evaluate the performance of theoretical models. The three presented sensors incorporate magnets of spherical and/or cylindrical geometry and can be operated at adjustable frequencies within the range of $\SIrange{61.9}{307.3}{\Hz}$. 
Our proposed method to obtain the natural frequency demonstrates notable robustness to variations in the initial deflection amplitudes and quality factors resulting in statistical errors on the mean smaller than $\SI{0.05}{\percent}$. We find that the distance-frequency relationship is well described by an adapted dipole model accounting for material and manufacturing uncertainties. 
Their combined effect can be compensated by an adjustment of a single parameter which drives the median model deviation generally below $\SI{0.2}{\%}$.
Our depicted methods and results are important for the design and calibration process of new sensor types utilizing the MMR technique.
\end{abstract}

\end{minipage}}}

\maketitle

\input{acronyms}

\section{Introduction}
Resonant systems are widely used in various technical contexts due to their distinct capability to efficiently absorb, store, and release energy at well-defined characteristic frequencies. In a suitable design, they can serve as a key component for the interconversion of different energy forms. In particular, the incorporation of permanent magnets in mechanically oscillating structures enables the coupling of a magnetic field to the mechanical motion which can be technically exploited. This method has facilitated the development process of multiple magneto-mechanical device and material concepts across diverse application settings.

Corresponding architectures are established within the field of \ac{MEMS}, e.g. in the form of wireless low-frequency power receivers~\cite{Challa_2012,Truong_2018,du2018,garraud2018microfabricated,halim2021,halim2022}, resonant energy harvesters~\cite{Amirtharajah1998,Makoto_Mizuno_2003,kulah2008,HALIM2018walking} or rate-integrating gyroscopes~\cite{pai2014}. Furthermore, similar arrangements are considered in the context of metamaterials made of magnetically coupled and/or tuned unit resonator cells~\cite{serra2018,wang2018,moleron2019,palermo2019,grinbergmagnetostatic2019,grinberg_robust_2020} or in the generation of \ac{ULF} magnetic fields in underwater communication systems~\cite{fawole2017,m_n_magnetic_2019,Fereidoony2022,Thanalakshme2022,jing2023}. The latter are often summarized as \acp{MMT} and are based on bearing-mounted permanent magnets~\cite{kanj2022,li2025} allowing for a torsional oscillation. 

Only very recently a further group of resonant devices gained attention that consists of passive, wireless sensors and tracking tools. Miniature mechanical resonators containing permanent magnets have shown remarkable potential with respect to signal strength and scalability. These are either based on a flexible cantilever~\cite{fischer_magneto-oscillatory_2024,fischermagnetfield2024} or on a purely magnetic restoring torque~\cite{gleich2023miniature}. Sensors of these types have many advantages over their wired counterparts in applications such as condition and structural monitoring, process control, and healthcare~\cite{bogue2010wireless,he2023,huang2016}. 
In particular, they can play a crucial role for the accurate determination of the position and orientation of medical instruments which is essential for procedures such as surgery, endoscopy, and vascular interventions~\cite{ramadani_survey_2022,kaesmacher2018reasons,mont2020unfavorable}. Also, the performance evaluation of (bio-)chemical reactions in industrial reactors can benefit from the precise knowledge of spatially distributed process parameters~\cite{mears1971tests, shabanian2017effects}. The potential applicability of the \ac{MMR} platform proposed by Gleich et al.~\cite{gleich2023miniature} in such scenarios motivates the investigation of this study.

The setup from~\cite{gleich2023miniature} has the advantage to allow the determination of all six spatial degrees of freedom of a sensor (tracking) and the simultaneous measurement of additional physical parameters such as temperature and 
\begin{figure}[t]
	\centering \includegraphics[width=\columnwidth]{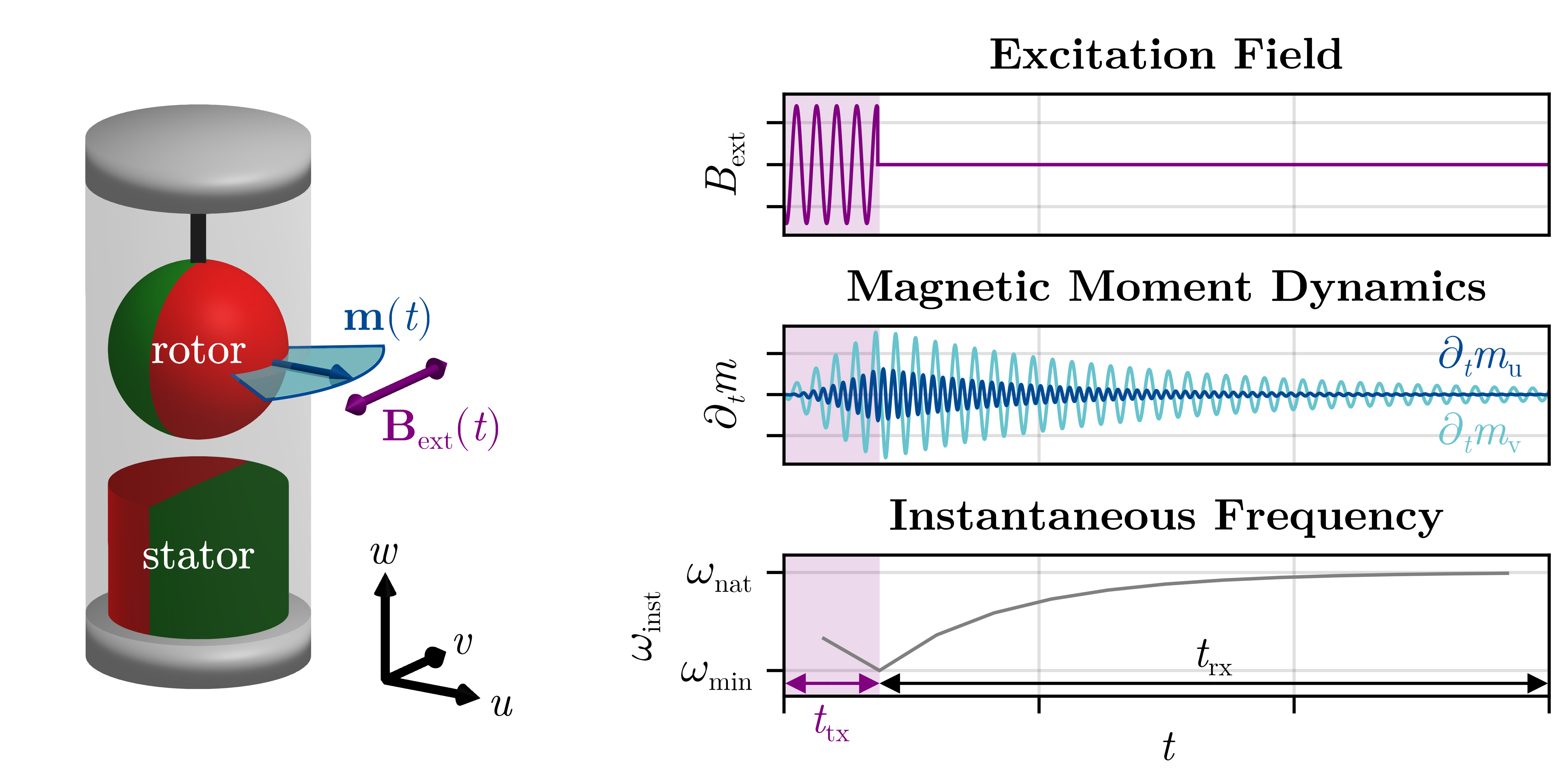}
	\caption{Left: Fundamental components of an \ac{MMR}. The filament keeps the rotor at a defined distance to the stator. An external magnetic field $B_\text{ext}(t)$ can deflect the rotor's magnetic moment $\mathbf{m}(t)$ from its equilibrium alignment.
    Right: Typical \ac{MMR} measurement frame consisting of excitation ($t_\text{tx}$, purple shading) and receive window ($t_\text{rx}$).  The components $\partial_t m_{u,v}(t)$ are proportional to the induction signal in a coil. 
    Note that due to the projection, we find the torsional frequency only in $v$-direction (light blue) but twice of it in $u$-direction (dark blue).
    For oscillations with a high quality factor, the instantaneous angular frequency $\omega_\text{inst}$ of this non-linear oscillator approaches the natural angular frequency $\omega_\text{nat}$ in the limit of vanishing deflection amplitude.}
	\label{img:basicPrinciple}
\end{figure}
pressure (sensing). As outlined in Fig.~\ref{img:basicPrinciple}, the design of the \ac{MMR} sensor includes two permanent magnets with antiparallel magnetic moments. The stator magnet is attached to the housing, whereas the rotor is suspended by a filament, able to rotate around the filament axis. The fundamental principle of the \ac{MMR} is similar to a torsional pendulum where the rotor performs damped oscillations around the equilibrium alignment after initial deflection by an external magnetic field.
The necessary restoring torque is supplied by the magnetic interaction with the stator. In analogy to \ac{LC} passive wireless sensors~\cite{collins1967miniature, huang2016}, the signal of the \ac{MMR} sensor can be excited and detected via induction.

We focus this work on the sensing aspect of \acp{MMR}. To achieve sensing capabilities, an external parameter like the environment pressure or temperature needs to be coupled to the oscillator. In the simplest approach, this can be realized by changing the magnet distance with e.g.~a compressible housing~\cite{gleich2023miniature,merbach2025pressure} or thermally deforming filament~\cite{gleich2023miniature}. This results in a parameter-dependent alteration of the restoring torque and thus of the measurable frequency. 
We note that non-linear oscillators show an additional dynamic coupling between the deflection angle and the instantaneous oscillation frequency, as illustrated in Fig.~\ref{img:basicPrinciple}. To guarantee well-defined \ac{MMR} sensing, we introduce the deflection angle-independent natural frequency. Oscillators with high quality factor converge towards this frequency for infinitesimally small deflection amplitudes.

In this paper, we investigate the sensor natural frequency response for varying rotor-stator distances and separate our analysis from any specific external physical quantity. This makes our results applicable for sensors independent of the implemented method that enables a distance change. Our measurements are performed on \acp{MMR}, composed of magnetic spheres and cylinders, specially engineered to cover a wide range of magnet-to-magnet distances. We compare these data sets with the frequency prediction of a fully determined dipole model and phenomenological models to gain an understanding of underlying deviations. The findings from this paper are a generalized and methodically substantially optimized extension of preliminary work published as a conference proceedings paper~\cite{knopp2024emperical}.

\section{Materials and Methods}

\subsection{Sensor Design}

\renewcommand{\arraystretch}{1.2}
\begin{table}[b]
\caption{\label{tab:magnets}
Manufacturer specifications of the permanent magnets utilized for \ac{MMR} construction.}
\setlength{\tabcolsep}{3pt}
\begin{tabular}{|>{\centering\arraybackslash}p{1.5cm}|>{\centering\arraybackslash}p{0.8cm}|>{\centering\arraybackslash}p{2cm}|>{\centering\arraybackslash}p{2.2cm}|c|}
\hline
Geometry&
Grade&
Remanence&
Diameter\,$|$\,Height&
Mass \\
\hline
Spherical  & N40 & $\SIrange{1.26}{1.29}{\tesla}$ & $\SI{4}{\mm}$\,$|$\, -- & $\SI{0.35}{\g}$\\
Cylindrical  & N35 & $\SIrange{1.17}{1.24}{\tesla}$ & $\SI{4}{\mm}$\,$|$\,$\SI{4}{\mm}$ & $\SI{0.38}{\g}$\\
\hline
\multicolumn{5}{p{245pt}}{The magnets of spherical or cylindrical geometry have a similar footprint but differ in their stated degree of magnetization.}\\
\end{tabular}
\end{table}
\renewcommand{\arraystretch}{1.0}

In this study, we compare three \acp{MMR} that differ in the choice of the magnet geometry. For the permanent magnets, we select from either spherical (P/N 1752, EarthMag GmbH, Germany), or cylindrical (P/N 20451, MAGSY GmbH, Germany) neodymium-iron-boron (NdFeB) magnets with uniform diametral magnetization. Their detailed material properties are summarized in Table~\ref{tab:magnets}.
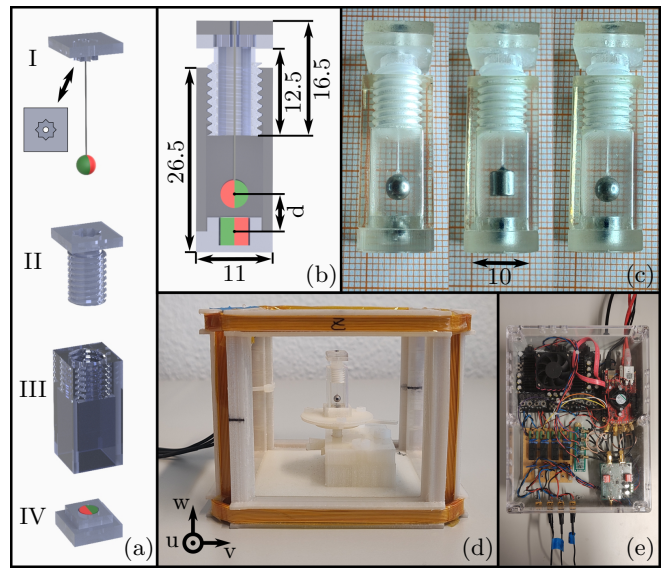
\begin{figure}[t]
\begin{tikzpicture}
   \node[anchor=south, inner sep=0] (image) at (0,0){\includegraphics[width=\columnwidth]{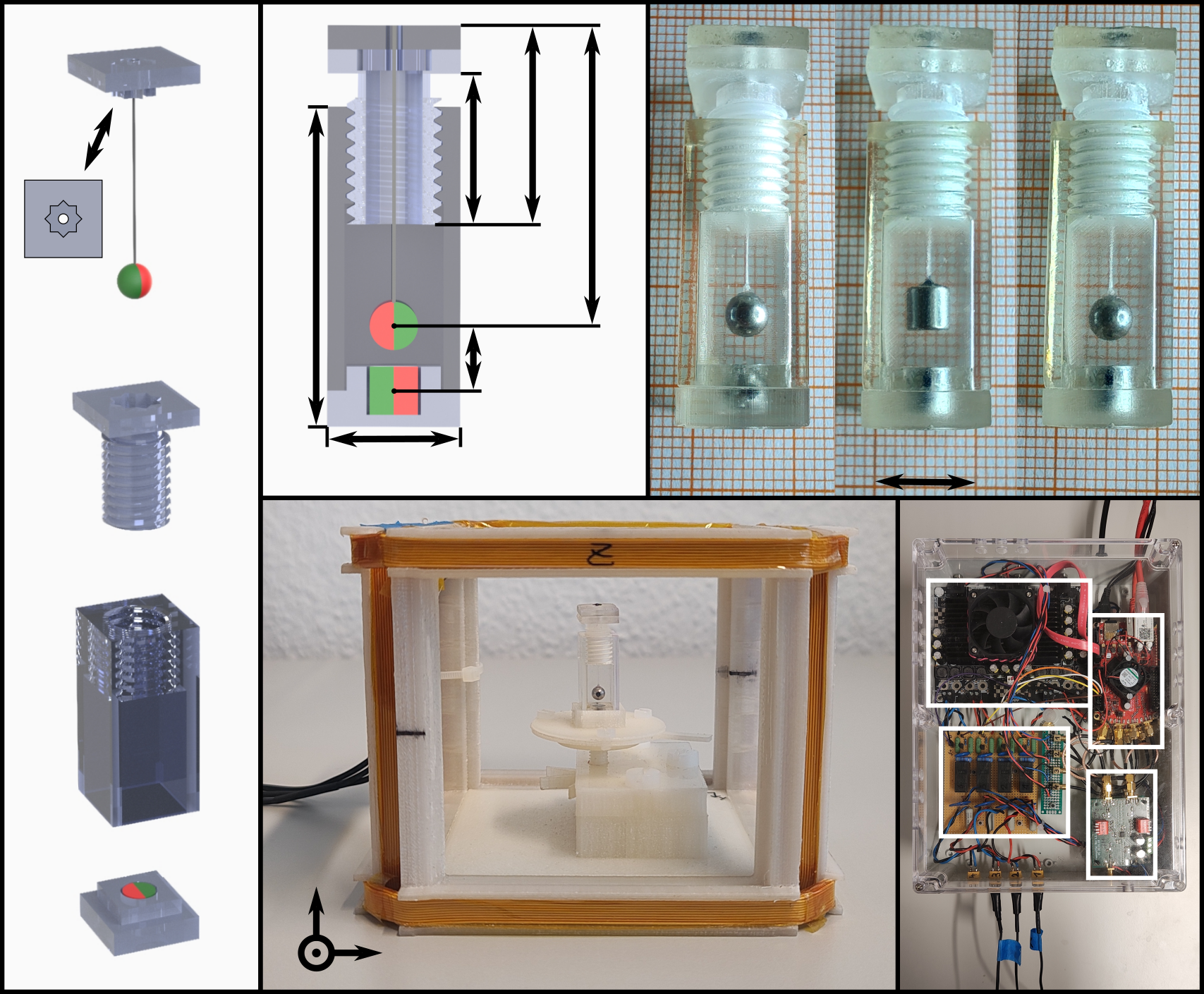}};

   \node[anchor = north, inner sep = 0] at (-4.13,0.79) {(4)};
   \node[anchor = north, inner sep = 0] at (-4.13,2.45) {(3)};
   \node[anchor = north, inner sep = 0] at (-4.13,4.15) {(2)};
   \node[anchor = north, inner sep = 0] at (-4.13,6.75) {(1)};

   \node[anchor = north, inner sep = 0, rotate = 90] at (-2.42,5.37) {26.5 mm};
   \node[anchor = north, inner sep = 0] at (-1.545,3.98) {11 mm}; 
   \node[anchor = north, inner sep = 0, rotate = 90] at (-0.864,6.32) {12.5 mm};
   \node[anchor = north, inner sep = 0, rotate = 90] at (-0.43,6.36) {16.5 mm};
   \node[anchor = north, inner sep = 0, rotate = 90] at (-0.02,6.02) {25 mm};
   \node[anchor = north, inner sep = 0, rotate = 90] at (-0.86,4.67) {d};
   
   \node[anchor = north, inner sep = 0] at (2.35,4.115) {10 mm};

   \node[anchor = north, inner sep = 0] at (-2.355,0.3875) {u};
   \node[anchor = north, inner sep = 0] at (-1.58,0.25) {v};
   \node[anchor = north, inner sep = 0] at (-2.23,0.89) {w};

    \node[anchor = south, inner sep = 0] at (-2.75,0.075) {(a)};
    \node[anchor = south, inner sep = 0] at (0.09,3.7) {(b)};
    \node[anchor = south, inner sep = 0] at (4.165,3.7) {(c)};
    \node[anchor = south, inner sep = 0] at (1.92,0.075) {(d)};
    \node[anchor = south, inner sep = 0] at (4.165,0.075) {(e)};

    \node[anchor = south, inner sep = 0 , text = white] at (3.0,3.08) {(2)};
    \node[anchor = south, inner sep = 0 , text = white] at (3.87,2.83) {(1)};
    \node[anchor = south, inner sep = 0 , text = white] at (3.82,0.495) {(4)};
    \node[anchor = south, inner sep = 0 , text = white] at (2.96,0.80) {(3)};

\end{tikzpicture}
\caption{Components of the measurement setup. (a) Exploded-view drawing showing the four-part concept of the adjustable 3D-printed \ac{MMR} housing. It consists of the filament anchor (1) with an octagram-shaped stamp (inset), the attached filament and rotor magnet, the adjustment screw (2) with M8 thread, the carrier element (3) and the stator mount (4). (b) Cross section of the housing. By rotation of the screw, the center-to-center distance $d$ between rotor and stator can be changed ($\Delta d = \SI{1.25}{\mm}$ per full rotation). Simultaneous raising of the anchor prevents filament twisting. (c) Images of the \mbox{Sphere-,} Cylinder- and Hybrid-\ac{MMR} (from left to right). (d) Three-axis coil arrangement aligned with the \ac{MMR} coordinate system from Fig.~\ref{img:basicPrinciple} and vertical translation mount for sensor centering. (e) Image of the three-channel transmit-receive chain electronics~\cite{mohn_low-cost_2025}. The main components for generating the excitation signal and for induction signal processing are the combined ADC/DAC (1), the transmit amplifier (2), the TxRx switches (3) and the receive amplifier (4). }
\label{img:Setup}
\end{figure}
The first \text{MMR} consists of two spherical magnets, the second of two cylinders and the third is built with a spherical rotor and a cylindrical stator. In the following, we refer to these sensors as Sphere-, Cylinder- and Hybrid-\ac{MMR}, respectively (see Fig.~\ref{img:Setup}c for images). The specialized design of our sensor housings features a mechanism to precisely adjust the center-to-center distance $d$ between the magnets allowing us to measure an isolated magnet distance series on an otherwise unchanged system.

Our cuboidal, non-magnetic housing design is based on four 3D-printed components using a stereolithography process (Form 3B, Clear Resin V4, Formlabs GmbH, Germany). We detail on the construction in Fig.~\ref{img:Setup}a, b. The continuous adjustment screw (Fig.~\ref{img:Setup}a, 2) is the central element for setting the distances $d$. Its lower end can be connected to the carrier element (Fig.~\ref{img:Setup}a, 3) via a machined M8 thread that allows a translational motion of $\Delta d = \SI{1.25}{\mm}$ per full rotation. To enhance transparency of the housing, the outer facets of the carrier element are polished and finished with clear lacquer. The sub-mm thin filament is made of multiple strands of ultra-high-molecular-weight polyethylene (UHMWPE). With instant adhesive gel (Loctite 454, Henkel AG \& Co. KGaA, Germany), it is fixed only to the rotor magnet and inside a through hole of the filament anchor (Fig.~\ref{img:Setup}a, 1) which, as such, forms the suspension plane. Finally, the stator mount (Fig.~\ref{img:Setup}a, 4) completes the housing.

To avoid a changing contribution of the suspension to the restoring torque when the distance is adjusted, our design minimizes filament twisting by having no permanent connection with the rotatable adjustment screw. Instead, the anchor (along with the filament) is detachable and solely the magnetic attraction of rotor and stator hold the components in place. Hereby, the lower part of the anchor, which forms an octagram-shaped stamp, fits precisely into the corresponding cut-out of the adjustment screw. This design prevents slipping of the anchor during an oscillation but allows to separate the filament and adjustment screw during a distance change. Defined by the symmetry of the stamp, one eighth of a full rotation represents our smallest distance discretization which does not lead to an increased state of the filament torsion.
 
\subsection{Excitation and Signal Acquisition}
Our \ac{MMR} measurement sequence consists of a number of repeated frames with identical time span. In accordance with Fig.~\ref{img:basicPrinciple}, each frame is again divided into an excitation window of time $t_\text{tx}$, and receive window $t_\text{rx}$. We implement a \ac{TxRx} system based on the switching concept demonstrated in~\cite{mohn_low-cost_2025} employing a power amplifier. For this, we use synchronized electro-mechanical relays to switch between dedicated excitation and receive electronics and to ensure full switching of all the electronic parts, we wait a short time $t_\text{sw}$ during the transitioning of the windows.
The central element for the interaction with the sensor is a cube consisting of three orthogonal pairs of square-shaped coils each acting as a separate inductive excitation and acquisition channel. The corresponding components are presented in Fig.~\ref{img:Setup}d, e.

Each of the two lateral coil pairs feature an edge length and a distance of $\SI{10}{\cm}$ whereas the vertical pair has an edge length of $\SI{12.8}{\cm}$ and a distance of $\SI{8}{\cm}$. Each individual coil is made of $40$ turns of thin litz wire consisting of $500$ individual copper strands with a diameter of $\SI{20}{\um}$. Biot-Savart simulations show that within a volume element of $\SI{1}{\cm^3}$ around the center of the setup, the expected magnetic field inhomogeneity is less than $\SI{1.5}{\%}$ and field strengths exceeding $B_\text{ext}=\SI{40}{\micro \tesla}$ can be realized. 

We control the measurement sequence by a computer connected to a system-on-a-chip (STEMlab 125-14, Red Pitaya d.o.o., Slovenia) that acts as our multi-channel \ac{DAC} during the excitation window and as \ac{ADC} during the receive window~\cite{hackelberg2022flexible}. Both the DAC and the ADC operate at a sampling rate of $\SI{125}{\mega \sample / \s}$ with a bit depth of $14$ bits. To improve the signal quality, the ADC signal is decimated by a factor of $2048$, resulting in an effective sampling rate of $\sim \SI{61}{\kilo \sample / \s}$. Given these sampling constraints, the \ac{TxRx} timings ($t_\text{tx}, t_\text{sw}, t_\text{rx}$) can be freely tuned.
During the excitation window, we deflect the \ac{MMR} rotor from its equilibrium alignment. To this end, a sinusoidal \ac{DAC} voltage output is amplified by a four-channel class-D amplifier (Sure Electronics Co., Ltd., Malaysia) which supplies the coils with current via a low-impedance circuit. The system allows for dynamic adaptions of the excitation signal amplitude, frequency and phase of each channel. In our setup, we find a linear behavior between \ac{DAC} output and coil current amplitude for frequencies up to $\sim 400\,\text{Hz}$.
In the receive window, we measure the induction signals created by the oscillating magnetic moment of the \ac{MMR} rotor. Hence, the relays switch the coils into a high-impedance circuit. For analog-to-digital conversion, we amplify the acquired voltage with a custom operational amplifier by a factor of $\sim 100$.

Currently, we do not use a control ensuring an in-phase relation between residual motion of the oscillator at the end of a frame and the newly applied excitation signal of the next frame. However, in parallel to signal acquisition, we perform a real-time \ac{FFT} for an active excitation frequency control. Starting with an initial guess for the first frame, the Fourier spectrum determines the set frequency for the excitation window of each immediately following frame. 
The corresponding value is not to be confused with the natural frequency which is independent on the specific rotor deflection angle and derived solely during post-processing (see \mbox{section~\ref{sub:natural}}).

\subsection{Experimental Execution}
The primary objective of this paper is to study the relation between the \ac{MMR} rotor-stator distance and its natural oscillation frequency.
For each of the three \acp{MMR}, we start the measurement process by adjusting a center-to-center distance of $d=\SI{10}{\mm}$. With support of the cuboidal format and transparency of the housing, we can verify the magnet distance on a millimeter-scaled graph paper, analogues to Fig.~\ref{img:Setup}c.
Starting from that, any subsequent distance is obtained from the (fractional) number of rotations of the adjustment screw and its pitch, down to the minimum value of $d = \SI{4.2}{\mm}$. Before each rotation, we lift the anchor, adhered filament and rotor to prevent twisting of the filament. We perform all of our measurements at room temperature, such that there is no relevant influence of thermal expansions in the sensor components on the adjusted distance.

As shown in Fig.~\ref{img:Setup}d, we mount the \ac{MMR} upright into the cube containing the induction coil pairs by support of a 3D-printed translation- and rotation-mount (Ultimaker 3, PLA, Freeform4U GmbH, Germany). With that, we ensure that the rotor is vertically aligned with the center of the homogeneous region for each measured distance. We also rotate the \ac{MMR} to superimpose coil axes and \ac{MMR} coordinate system ($u,v,w$) by utilizing the projection characteristics from Fig.~\ref{img:basicPrinciple}.
To that end, we adjust the orientation of the \ac{MMR} such that after excitation we observe the torsional frequency dominantly in one lateral receive channel, and twice its value dominantly in the other. Consequently, the individual channels will be mainly sensitive to respective changes in $m_v$ or $m_u$. During alignment, the vertical coil pair acts as a verification channel as it does not show any induction if the sensor is positioned correctly. In the measurement sequence, we apply the excitation field only in $v$-direction which is ideal for efficient \ac{MMR} stimulation whereas the reception occurs in both lateral directions leading to a two-dimensional vector-valued receive signal $\pmb{u}(t)$.

To ensure statistical reliability, $17$ measurement repetitions are performed for each \ac{MMR} and distance, implemented as a measurement sequence of that number of subsequent frames. We fix $t_\text{sw} = \SI{20}{\ms}$ while the excitation times and field amplitudes are chosen for each individual sequence between $t_\text{tx}=\SIrange{0.5}{1.0}{\s}$ and $B_\text{ext}=\SIrange{20}{30}{\micro \tesla}$, respectively, to maximize the initial induction signal amplitude. As the sensors' damping timescales differ in dependence of $d$, we choose receive windows between $\SIrange{2}{30}{\s}$, to capture nearly the full decay of the oscillations up to the point where the signals get noise-dominated.

\subsection{Equation of Motion}\label{sub:modelfit}
We consider an \ac{MMR} model where both the rotor and stator are represented as magnetic point dipoles with magnetic moments $m_\text{r}$ and $m_\text{s}$, respectively. In that case, the instantaneous magnetic restoring torque $\tau = m_r B_0 \sin \varphi$ on a rotor that is deflected out of its equilibrium alignment depends on the magnetic flux density at the location of the rotor $B_0(m_s)$ produced by the stator and the deflection angle $\varphi$~\cite{gleich2023miniature}. We note that in designs of other magneto-mechanical oscillators, the suspension of the rotor provides an additional mechanical restoring torque~\cite{jing2023,li2025}.
However for our sensors, we neglect any residual contribution of the thin filament and consider the torque as dominated by the magnetic interaction. Finally, we account for frictional losses such that the corresponding equation of motion for $\varphi$ in free oscillation is equivalent to that of a damped gravitational pendulum and given by
\begin{equation} \label{eq:ode}
\ddot{\varphi} + \frac{\omega_\text{nat}}{Q} \dot{\varphi} + \omega_\text{nat}^2 \sin{\varphi} = 0
\end{equation}
where $\omega_\text{nat}$ is the natural angular frequency of the resonator and $Q$ is the quality factor affecting the relaxation time \mbox{$\tau = 2Q/\omega_\text{nat}$.}

In the small-angle approximation ($\sin\varphi \approx \varphi$), Eq.~\eqref{eq:ode} becomes the equation of motion of a damped harmonic oscillator, which has the explicit solution
\begin{equation}
\varphi(t) = \varphi_\text{max}\exp\left(-t/\tau\right) \sin \left(\omega_\text{inst}\,t+\psi_0 \right)
\end{equation}
where $\varphi_\text{max}$ is the maximum (initial) deflection angle and $\psi_0$ is an initial phase at $t=0$. Only in the case of a high-$Q$ oscillation, the instantaneous angular frequency
\begin{equation}\label{eq:instafreq}
\omega_\text{inst} = \sqrt{1- \left( \frac{1}{2Q} \right)^{2}} \,\omega_\text{nat}
\end{equation}
approaches the natural angular frequency, as illustrated in Fig.~\ref{img:basicPrinciple}. In other words, $\omega_\text{nat}$ corresponds to the angular frequency at which an undamped harmonic oscillator would oscillate.

\subsection{Natural Frequency Estimation}\label{sub:natural}
To determine the natural oscillation frequency of an \ac{MMR}, we perform a direct optimization on the physical model from the equation of motion given by Eq.~\eqref{eq:ode}. Compared to an isolated analysis of the instantaneous frequency in a late damped-out state, as possible with Eq.~\eqref{eq:instafreq}, our method has the advantage to take the full acquired induction voltage signals $\pmb{u}(t)$ as an input. Consequently, our estimation performance is expected to be independent of $Q$ and not sensitive to varying initial signal amplitudes and corresponding fluctuations in the onset of the small-angle approximation. 

Since we do not directly measure $\varphi$, we need to relate the quantity with the received voltage signal of each frame. This relation can be modeled by 
\begin{equation}\label{eq:signal}
\pmb{u}(t) = \pmb{\sigma}_v \partial_t  \sin(\varphi(t)) + \pmb{\sigma}_u \partial_t  \cos(\varphi(t))
\end{equation}
where $\pmb{\sigma}_v$ and $\pmb{\sigma}_u$ are amplitude vectors, inherently dependent on $m_r$ and the coil sensitivities. The two terms are explained from the time-dependent magnetic moment projections in either $v$- or $u$-direction, for which a general coil has distinct sensitivities.
In our alignment, each lateral receive coil pair is predominantly sensitive to one respective component of these two.
For each set of the parameters $\pmb{\sigma}_v$, $\pmb{\sigma}_u$, $\omega_\text{nat}$, $Q$ and the initial $\varphi_\text{max}$ and $\psi_0$, we can determine the corresponding voltage time-signal following from the Eq.~\eqref{eq:ode} and \eqref{eq:signal}. For each frame, we optimize all these parameters to fit the measurement data considering a \ac{NLLS} approach by means of the \ac{LMA}~\cite{levenberg1944method}.

\subsection{Frequency Models}
\renewcommand{\arraystretch}{1.2}
\begin{table}[b]
\caption{\label{tab:parameters}
Magnetic and mechanic properties of the rotor and stator.}
\setlength{\tabcolsep}{3pt}
\begin{tabular}{|>{\centering\arraybackslash}p{1.1cm}|>{\centering\arraybackslash}p{2.4cm}|c|c|}
\hline
Symbol&Quantity&\makecell{Magnitude\\spherical magnet} & \makecell{Magnitude\\cylindrical magnet} \\
\hline
$m_r$, $m_s$  & \makecell{rotor/stator\\magnetic moment} & $\SI{3.4e-2}{\ampere \,\meter^2}$ & $\SI{4.8e-2}{\ampere \,\meter^2}$\\
$I$  & \makecell{rotor\\moment of inertia} & $\SI{5.6e-10}{\kg\,\m^2}$ & $\SI{7.6e-10}{\kg \,\m^2}$\\
\hline
\multicolumn{4}{p{245pt}}{The stated magnitudes are used to compute the frequency prediction of the physics-motivated dipole model A.}\\
\end{tabular}
\end{table}
\renewcommand{\arraystretch}{1.0}

To derive an expression for the natural angular frequency in dependence on material parameters, we assign to the modeled point dipoles of rotor and stator (located at their respective center-of-mass) the mechanical properties resulting from their actual extended shape.
To comply with the definition for $\omega_\text{nat}$ of section~\ref{sub:modelfit}, we consider an undamped harmonic (i.e. small-angle approximated) torsional motion with the torsion constant $D = m_r B_0$ and the rotor's moment of inertia $I$. 
In this case, we can write $\omega_\text{nat} = \sqrt{D/I}$~\cite{gleich2023miniature}.
For the dipole field produced by the stator, we have $B_0(d) =  \mu_0m_\text{s} /(4\pi d^3)$ where $d$ is the distance between both magnet centers and $\mu_0 $ is the vacuum permeability. Thus, the angular natural frequency of this dipole model, labeled as model A, follows as
\begin{equation}\label{eq:res}
    \text{Model A}: \quad \omega_\text{nat}^\text{A}(d) = \alpha^\text{A}\, d^{-3/2}
\end{equation}
with
\begin{equation}\label{eq:prefactorA}
    \alpha^\text{A}:= \sqrt{\frac{m_\text{r} m_\text{s} \mu_0}{4\pi I}}.
\end{equation}
The material factor $\alpha^A$ depends only on the magnetic and mechanical properties of the sensor components. To obtain this quantity, we consider rotor and stator as geometric objects with uniform magnetization and mass distribution, and take the manufacturer specifications from table~\ref{tab:magnets} as a reference. Thereby, we use the center value of the stated remanence range to calculate the magnetic moment of the respective magnet. For the moment of inertia, we assume an ideal rotation of the suspended magnet around the axis of rotational symmetry (passing through the center-of-mass) for either a spherical or cylindrical shape with corresponding size and mass. The obtained values for $m_r$, $m_s$ and $I$ are summarized in table~\ref{tab:parameters}.

The frequency prediction of model A is based on various assumptions regarding the construction of our sensors. To allow for a systematic analysis of the contribution of inaccuracies, we derive from model A two additional phenomenological models. The first uncertainty comes from the precision in the knowledge of material and construction parameters. Both, the obtained rotor and stator magnetization, and the moment of inertia rely on the manufacturer specifications and idealizations in the determination. These parameters contribute to the material factor $\alpha^\text{A}$ of model A which is, therefore, chosen as a free parameter in model B. The second uncertainty lies in the validity of the dipole approximation to describe the real interaction of our spatially extended magnets. The dipole assumption is well established to describe the magnetic field of an isolated magnet at large separations~\cite{petruska2013dipole}, for ideal magnetic spheres even in the near-field~\cite{Jackson2014}. As our sensors are subject to magnet edge-to-edge distances on the order of the magnet size and below, we aim to experimentally quantify deviations from a dipole scaling due to the near-field coupling of rotor and stator. Consequently in model C, we allow for the additional adaptation of the polynomial degree of the spatial profile of $B_0$. 
In summary, these models are described by
\begin{align}
    \text{Model B}: \quad& \omega_\text{nat}^\text{B}(d,\alpha^B)  = \alpha^B\, d^{-3/2}, \\
    \text{Model C}: \quad& \omega_\text{nat}^\text{C}(d, \alpha^C, \gamma^C)  = \alpha^C\, d^{-\gamma^C/2}, 
\end{align}
with free parameters $\alpha^B$, $\alpha^C$ and $\gamma^C$.
These are optimized for each model on the measured frequency values employing a weighted \ac{NLLS} fit minimizing the sum of the squared relative residuals using the \ac{LMA}. We quantify the scale of adaption compared to the fully determined model A by computation of the relative parameter deviation $\Delta \alpha = (\alpha^{B,C}-\alpha^A) / \alpha^A$ and $\Delta \gamma = (\gamma^{C}-\gamma^A) / \gamma^A$ with $\gamma^A = 3$.

To evaluate the agreement between the established model $M \in \{A, B, C\}$ and the measurement data, we apply an analysis of the relative residuals defined by
\begin{align}\label{eqn:error}
\varepsilon^{\text{M}}(d) &= \frac{\omega_\text{nat}^\text{M}(d) -\omega_\text{nat}^\text{meas}(d)}{\omega_\text{nat}^\text{meas}(d)}
\end{align}
for each distance. Thereby, $\omega_\text{nat}^\text{meas}$ is the natural angular frequency obtained from averaging all frames from a single measurement sequence.

\section{Results}
\begin{figure}[t]
	\includegraphics[width=\columnwidth]{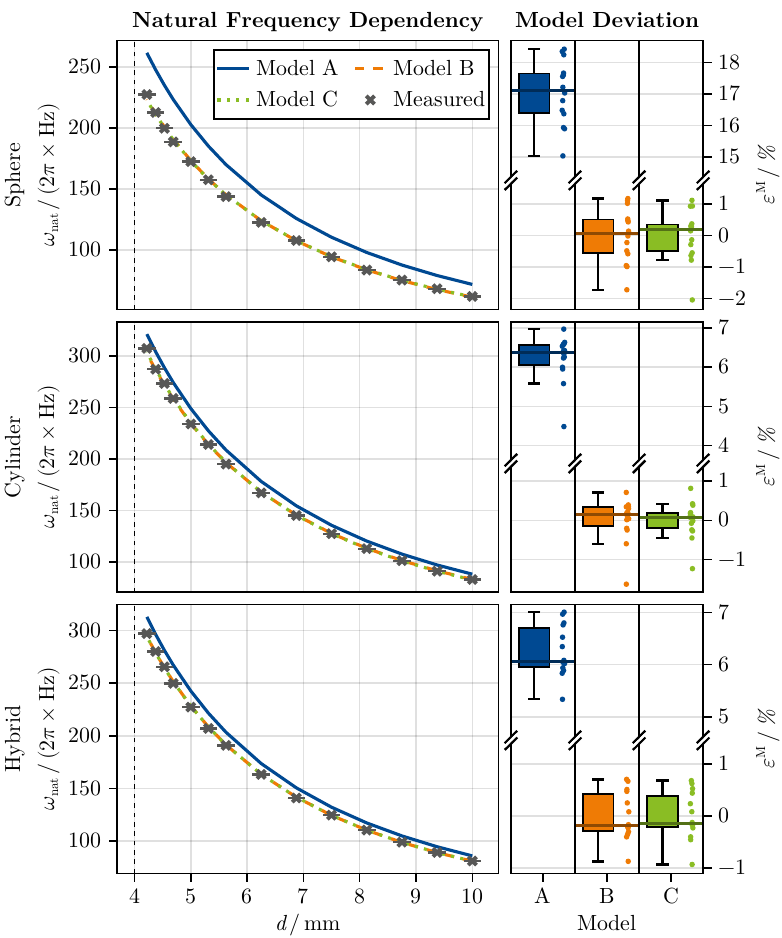}
	\caption{Left: Measured mean natural angular frequencies $\omega_\text{nat}$ in units of $\text{Hz}$ for different magnet center-to-center distances $d$ in comparison to predictions from models A, B, and C. We show the data sets for the sensors labeled as Sphere- (first row), Cylinder- (second row) and Hybrid-\ac{MMR} (last row). Each frequency value is obtained from averaging $17$ subsequent frames where we generally find statistical errors on the mean smaller than the marker size. The horizontal bars show the experimental uncertainty in the determination of $d$. The black dashed line marks the contact point between rotor and stator. Right: Box plots combining the model deviations $\varepsilon^{\text{M}}$ according to Eq.~\eqref{eqn:error} for all measured distances. Each dot represents a sequence-averaged data point, the box limits are defined by the lower and upper quartile, and the whisker expands the interquartile range by a maximum factor of $1.5$. The horizontal lines mark the median deviation. Note the break in the vertical axes for improved visibility.}
	\label{img:resultModelComparison}
\end{figure}
Our results for the relation between magnet distance and natural frequency are summarized in Fig.~\ref{img:resultModelComparison}. In the left part, we show the measured and modeled frequency values for each of the three constructed \acp{MMR}. Within a single measurement sequence, we observe large frame-to-frame variations in the initial deflection amplitude, with the largest range spanning $\varphi_\text{max} = \SIrange{9.6}{31.8}{\degree}$. 
Nonetheless, there is a strong reproducibility in the determination of the natural frequency yielding statistical errors on the mean generally smaller than $\SI{0.05}{\%}$. 
Depending on the measurement configuration, we find sequence-averaged quality factors ranging from $ Q = \SIrange{186}{14052}{}$. The quality factors show frame-to-frame fluctuations during a measurement sequence, where the maximum standard deviation of the mean lies at $\SI{16.2}{\%}$. We note that the relation between magnet distance and $Q$ does not exhibit a simple, universal scaling behavior in our measurements.

\renewcommand{\arraystretch}{1.2}
\begin{table}[b]
\caption{\label{tab:results} Comparison of model performances.}
\setlength{\tabcolsep}{4pt}
\begin{tabular}{|>{\centering\arraybackslash}p{1.0cm}|>{\centering\arraybackslash}p{1.1cm}|>{\centering\arraybackslash}p{0.95cm}>{\centering\arraybackslash}p{0.95cm}|>{\centering\arraybackslash}p{0.96cm}>{\centering\arraybackslash}p{0.96cm}c|}
\hline
\multirow{2}{*}{\ac{MMR}} &
Model A & \multicolumn{2}{c|}{Model B} & \multicolumn{3}{c|}{Model C}\\
 & $\tilde\varepsilon^A / \SI{}{\%}$ &  $\Delta\alpha / \SI{}{\%}$ & $\tilde\varepsilon^B / \SI{}{\%}$ & $\Delta\alpha / \SI{}{\%}$ & $\Delta\gamma / \SI{}{\%}$ & $\tilde\varepsilon^C / \SI{}{\%}$ \\
\hline
Sphere  & $\SI{17.1}{}$ & $\SI{-14.6}{}$ & $\SI{0.1}{}$ &$\SI{-10.8}{}$ & $\SI{-0.6}{}$ & $\SI{0.2}{}$ \\
Cylinder  & $\SI{6.4}{}$ & $\SI{-5.8}{}$ & $\SI{0.1}{}$ & $\SI{-10.7}{}$ & $\SI{0.7}{}$ & $\SI{0.1}{}$ \\
Hybrid  & $\SI{6.1}{}$ & $\SI{-5.9}{}$ & $\SI{-0.2}{}$ & $\SI{-5.1}{}$ & $\SI{-0.1}{}$ & $\SI{-0.1}{}$\\
\hline
\multicolumn{7}{p{245pt}}{For the three sensors, we present the median of the distance-combined model deviation distribution of Fig.~\ref{img:resultModelComparison}, denoted by $\tilde\varepsilon^M$. We also show the relative deviations $\Delta \alpha$ and $\Delta \gamma$ for the optimized free parameters in model B ($\alpha^B$) and C ($\alpha^C$, $\gamma^C$) compared to the values of model A.}
\end{tabular}
\end{table}
\renewcommand{\arraystretch}{1.0}
In general, the oscillation frequency is increasing with decreasing center-to-center distance for all investigated sensors. The measured frequencies span a range of $\SIrange{61.9}{227.6}{\Hz}$ for the Sphere-\ac{MMR}, $\SIrange{82.8}{307.3}{\Hz}$ for the Cylinder-\ac{MMR} and $\SIrange{81.0}{297.1}{\Hz}$ for the Hybrid-\ac{MMR}.
For each model and sensor, we show the statistics of the distance-combined $\varepsilon^{\text{M}}$ as box plots on the right side of Fig.~\ref{img:resultModelComparison}. In addition, we evaluate in table~\ref{tab:results} the degree of adaption of the free parameters in the phenomenological models compared to model A and specify the overall model performances. In comparison to our measurements, the dipole model A predicts higher frequencies for all sensors with the strongest discrepancy for the sensor with spherical rotor and stator. 
We observe that the material factors $\alpha^B$ are generally reduced compared to $\alpha^A$ with the largest percentage adjustment again for the Sphere-\ac{MMR}. In model B, the absolute median model deviations can be brought to a similar level below $\SI{0.2}{\%}$ for all three \acp{MMR} with a maximum interquartile range of $\SI{1.1}{\%}$.
The $\varepsilon^M$ from the least restrictive model C are comparable to these values. Regarding to the dipole value, the exponent $\gamma^C$ needs to be only slightly adjusted by not more than $\SI{0.7}{\%}$.

\section{Discussion}
The observed frame-variations in the initial deflection amplitude of our experiment are attributable to the absence of active excitation signal phase control. At the initiation of a frame, this can result in either constructive or destructive interference between the rotor's residual small-amplitude motion from the previous frame and the subsequently applied excitation signal. Consequently, the time at which the oscillator begins to effectively follow the external drive varies, resulting in non-uniform excitation processes. Although the implementation of a suitable phase forecasting algorithm may resolve this issue, the statistically insignificant errors observed highlight the inherent robustness of our employed methodology to determine the natural frequency against this limitation. Moreover, the determination process also tolerates the strong variability in the measured $Q$ factors. This pronounced insensitivity to the overall quality of an excitation is advantageous for real sensing applications since there is no need to achieve precisely constant rotor deflection angles or quality factors. We note that an analysis of the precise mechanism for the non-trivial scaling of the quality factor with the magnet distance must be addressed in specialized future studies that are outside the scope of this work.

Qualitatively, we observe for the measurements and models an increase in the natural frequency for smaller magnet distances resulting from the rising magnetic field strength and restoring torque on the rotor. However, we have a systematic discrepancy with the fully determined dipole model A towards overestimated frequency predictions with median model deviations exceeding $\SI{6}{\%}$. This trend is consistent with the findings from~\cite{jing2023}. In contrast, the models B and C both drive the model error close to zero. Compared to the dipole value, the fitting systematically reduces the material factors $\alpha^{B,C}$ for all sensors, peaking at $\SI{-14.6}{\%}$ for the Sphere-\ac{MMR} in model B. There is no substantial further improvement in the model performances $\tilde\varepsilon^{B,C}$ when transitioning to model C with additional free parameter. In summary, the presented model A is not sufficient to describe the observations, whereas model B outperforms model C because of its reduced complexity. We note that due to the data-driven nature of the proposed model, small (systematic) deviations with the measurements still persist and ultimately limit the estimation accuracy and sensor sensitivity. 

These findings highlight the importance of an accurate material factor while the observed minor optimizations in $\gamma^C$ (below $\SI{0.7}{\%}$) compared to the dipole value can be attributed to overfitting on the measurement data and have no physical significance. 
From the latter, we emphasize that there is no indication to assume a spatial magnetic field profile other than that of a dipole field to predict the natural frequency in our experiment. The frequency has a direct link to the effective torque acting on the rotor. 
While it is also theoretically described that the torque between magnetically interacting spheres exactly follows the expectations of a dipole approach~\cite{edwards2017interactions}, the literature lacks a similar analysis for cylinders. However, our results indicate a torque scaling behavior as from a dipole model which is independent on the magnet geometry in this work. Any residual offsets seem to be mainly absorbed into the adequate adaption of the material factor.
Our findings contrast to results from measurements performed on similar \ac{MMT} devices where smaller exponents (corresponding to $\gamma^C\sim 2$) have been observed~\cite{Thanalakshme2022,kanj2022}. Also simulations suggest a rather large deviation from a dipole model frequency scaling due to near-field effects in these systems~\cite{jing2023}. We attribute this to the larger aspect ratio in the rotor/stator geometry in the corresponding setups which leads to a more significant impact of the magnets' spatial extent compared to our study. We infer that the compact format of our magnets is not only advantageous for further miniaturization but also, correspondingly, simplifies the modeling of the sensor response even for the near-field magnet-to-magnet coupling. 

The single parameter $\alpha^B$ compensates non-specifically for uncertainties of the magnet magnetization and the rotor's moment of inertia. The first can result from manufacturing discrepancies of the stated remanence or volume while the latter among other things may be affected by e.g. a small relative tilt of the magnets or the exact amount, shape and position of glue at the bonding point of rotor and filament. We note that for all sensors, a simple assumed remanence shift inside the limits of table~\ref{tab:magnets} is not sufficient to push the error of model A to the same level as with model B. Furthermore, the stated irregularities during assembly will more likely increase $I$ (compared to model A) which, according to Eq.~\eqref{eq:res} and \eqref{eq:prefactorA}, leads to smaller natural frequencies in agreement with our measurements.
Moreover, as indicated by the model and parameter deviations listed in table~\ref{tab:results}, the Sphere-\ac{MMR} demonstrates the highest susceptibility to corresponding construction-related deficits in our experiment. This finding aligns with the increased complexity associated with the sensor manufacturing process for magnets with spherical symmetry. In particular, the mounting for the precise attachment of the filament to the rotor as well as the alignment of the stator magnet in its mount become more demanding for spheres. Notably, the impact of the latter is significant, as the model deviations for $\alpha^B$ are comparable for the Cylinder- and Hybrid-\ac{MMR} both of which share the cylindrical stator but differ in the geometry of their rotors. We remark that larger permanent magnets typically reduce the construction effort as well as the relative influence of material and manufacturing irregularities due to their increased mass and magnetic moments. These slightly modified conditions apply in the preliminary experimental study of~\cite{knopp2024emperical} whose more pronounced dipole model agreement is therefore  consistent with our findings.

\section{Conclusions and Outlook}
We performed a systematic experimental study investigating the dependency of the natural \ac{MMR} frequency on the rotor-stator distance for different magnet geometries with the same characteristic size of $\SI{4}{\mm}$.
Notably, we were able to show that present sensor manufacturing uncertainties can be readily accounted for by the adjustment of the single parameter $\alpha^B$ in a point dipole based model. Correspondingly, the median deviations between measurements and model predictions reach $\SI{0.1}{\percent}$ for our sensors made of either spherical or cylindrical magnets and $\SI{-0.2}{\percent}$ for a hybrid sensor. 

Application-oriented sensors designed to measure external parameters can be conceptualized on varying magnet distances~\cite{gleich2023miniature,merbach2025pressure}. As our proposed model B is valid independent on the mechanism for the distance change, it can be readily combined with e.g.~pressure deformation/thermal expansion models of the housing which enables general parameter–frequency models. This generalization would be valuable for the sensor design process by allowing an accurate frequency range prediction and its adaption to application-specific requirements without the need for a time-consuming precision analysis of the mechanical oscillator materials and components. Furthermore, our work suggests the transition to a few-point calibration approach in combination with a phenomenological model to reduce the calibration effort of \ac{MMR} sensors.
These findings represent an important step towards the fully quantitative sensing of e.g.~temperature, pressure or other parameters based on magneto-mechanical resonance.

\bibliographystyle{IEEEtran}
\bibliography{ref}
\begin{IEEEbiography}[{\includegraphics[width=1in,height=1.25in,clip,keepaspectratio]{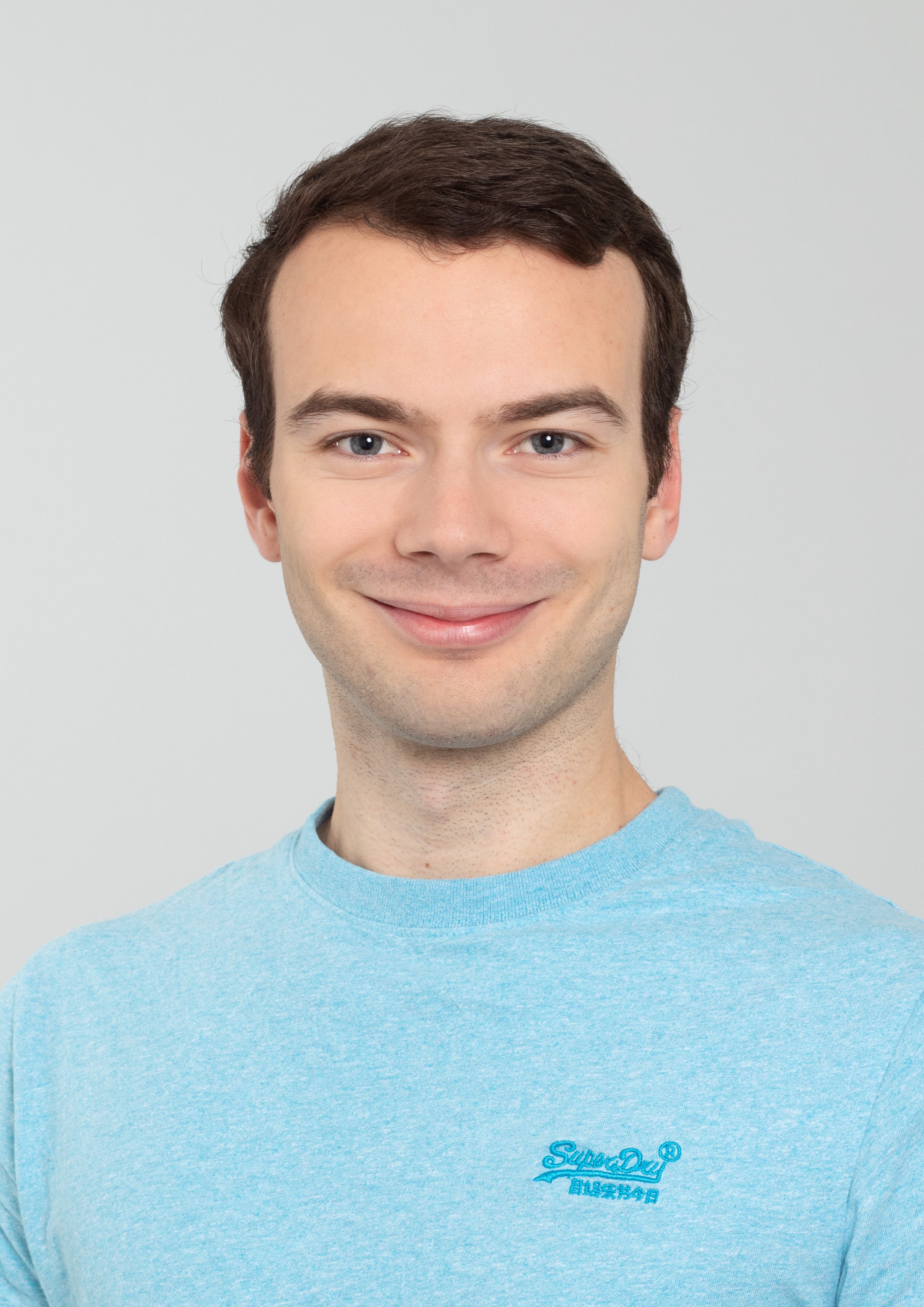}}]{Jonas Faltinath} received his B.Sc. and M.Sc. degree in physics from the University of Hamburg, Germany, in 2019 and 2022, respectively. Between 2022 and 2023, he performed a research stay at the Ecole Polytechnique Fédérale de Lausanne (EPFL), Switzerland. 

Currently, he is working as a research assistant at the University Medical Center Hamburg-Eppendorf, Germany and is pursuing the PhD degree at the Hamburg University of Technology, Germany. His current research interests include magneto-mechanical resonators, sensor conceptualization and development, and medical imaging techniques.
\end{IEEEbiography}

\begin{IEEEbiography}[{\includegraphics[width=1in,height=1.25in,clip,keepaspectratio]{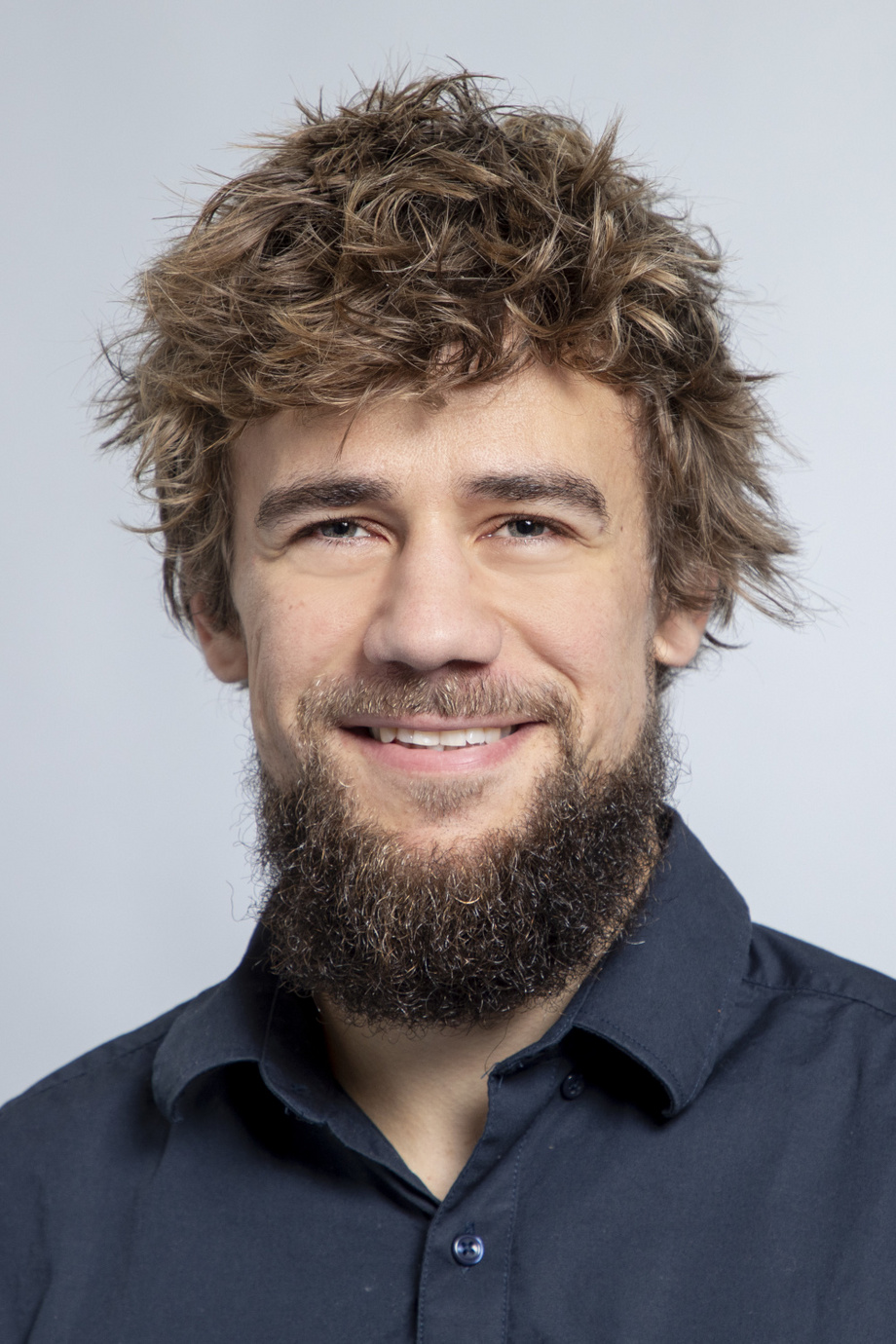}}]{Fabian Mohn} studied electrical engineering at the Hamburg University of Technology, Germany and received his master's degree in cooperation with the Philips Research Laboratories Hamburg, Germany in 2018. 

He joined the research group of Tobias Knopp for Biomedical Imaging at the University Medical Center Hamburg-Eppendorf and the Hamburg University of Technology, Germany in 2020 and received his PhD in 2024. His research interests include medical imaging and inductive sensors.
\end{IEEEbiography}

\begin{IEEEbiography}[{\includegraphics[width=1in,height=1.25in,clip,keepaspectratio]{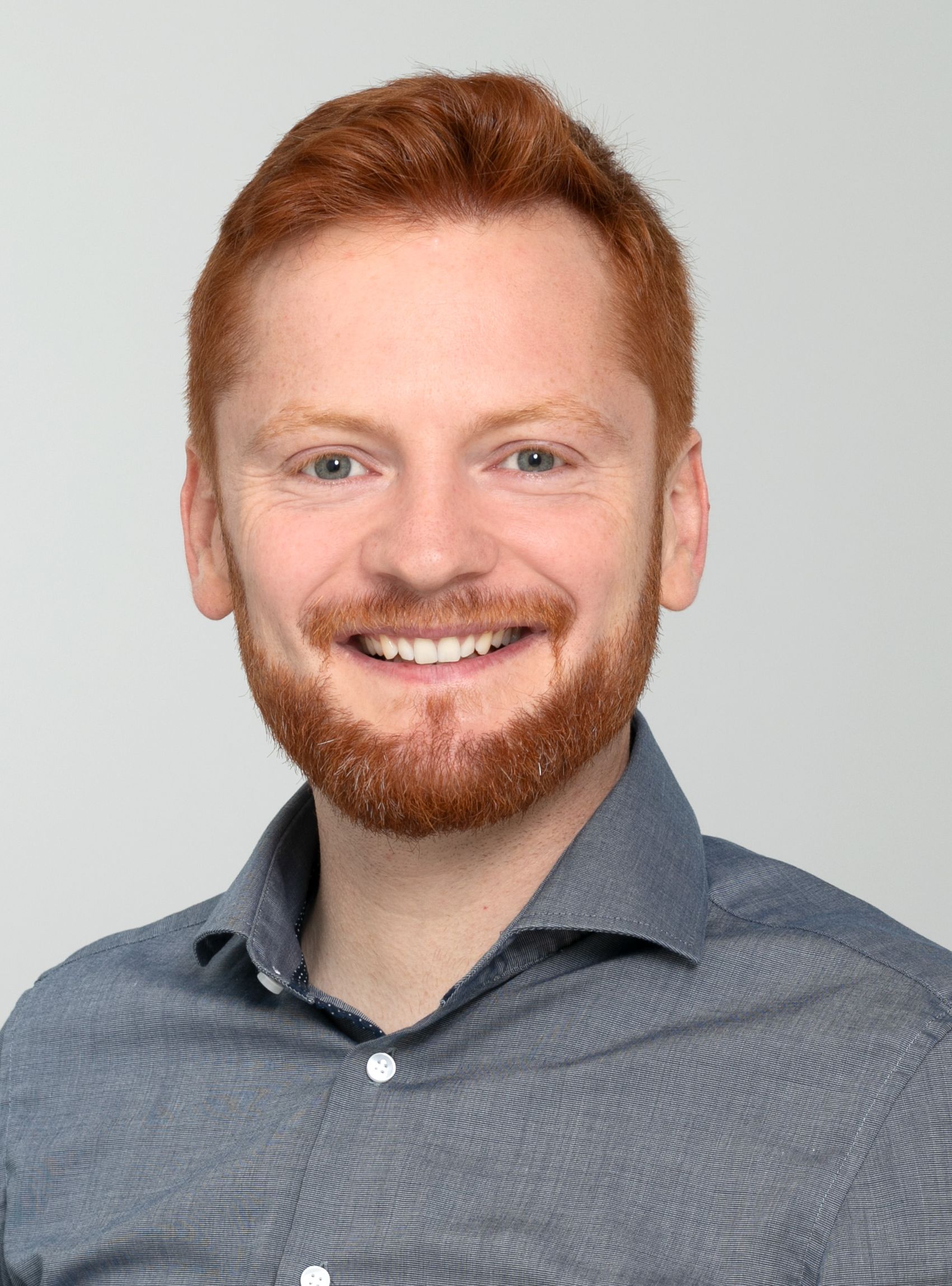}}]{Fynn Foerger} received his B.Sc. and M.Sc. degree in physics from the University of Hamburg, Germany, in 2016 and 2018, respectively.

Currently, he is working as a research assistant at the University Medical Center Hamburg-Eppendorf, Germany and is pursuing the PhD degree at the Hamburg University of Technology, Germany. His current research interests include medical imaging, magnetic field generation and characterization, as well as sensor design and conceptualization.
\end{IEEEbiography}

\begin{IEEEbiography}[{\includegraphics[width=1in,height=1.25in,clip,keepaspectratio]{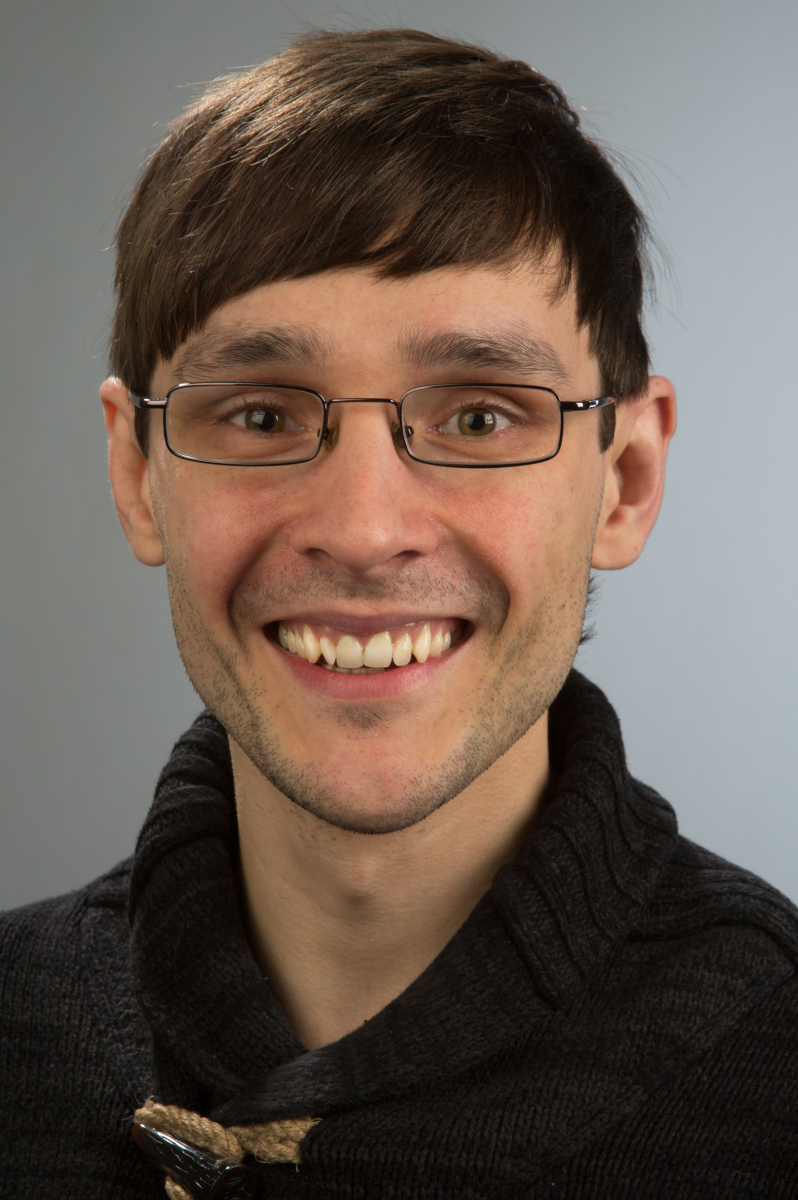}}]{Martin Möddel} received his physics Diploma in 2011 from the Leipzig University, Germany and his PhD in 2014 from the University of Siegen, Germany.

He is currently working as a postdoctoral researcher at the Institute of Biomedical Imaging which is jointly affiliated with the University Medical Center Hamburg-Eppendorf and the Hamburg University of Technology in Germany. His current research interests include inverse problems, particularly image reconstruction and artifact reduction, magnetic field generation and characterization, and signal processing, with applications in various areas of medical engineering.

\end{IEEEbiography}

\begin{IEEEbiography}[{\includegraphics[width=1in,height=1.25in,clip,keepaspectratio]{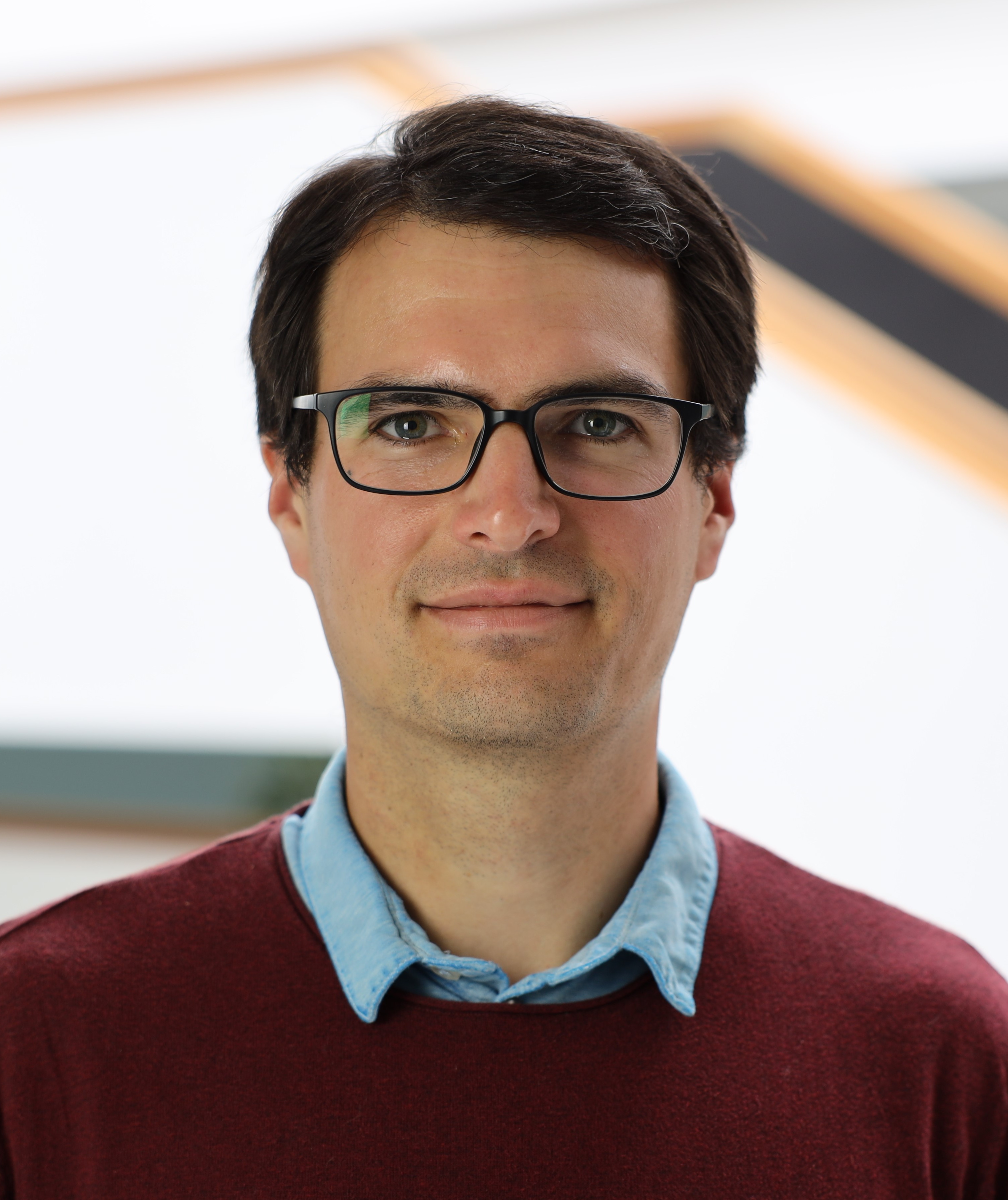}}]{Tobias Knopp} received his Diplom degree in computer science in 2007 and his PhD in 2010, both from the University of Lübeck, Germany.

He is currently a professor of Biomedical Imaging with a joint appointment at the University Medical Center Hamburg-Eppendorf and the Hamburg University of Technology, Germany. In addition, he heads the Department of Diagnostics at the Fraunhofer Research Institution for Individualized and Cell-Based Medical Engineering IMTE in Lübeck, Germany.
His current research interests include signal processing, inverse problems, tracking, sensing, and image reconstruction, with applications across various areas of medical engineering.
\end{IEEEbiography}

\end{document}

%% file: acronyms.tex
\begin{acronym}[ECU]

\acro{AWMPS}[AWMPS]{arbitrary waveform magnetic particle spectrometer}
\acro{ADC}[ADC]{analog-to-digital converter} 
\acro{AUC}[AUC]{area under the curve}

\acro{CT}[CT]{computed tomography}

\acro{DAC}[DAC]{digital-to-analog converter}
\acro{DFG}[DFG]{drive-field generator}
\acro{DFCs}[DFCs]{drive-field coils}
\acro{DF}[DF]{drive-field}
\acro{DA}[DA]{differential amplifier}
\acro{DLS}[DLS]{dynamic light scattering}
\acro{DTS}[DTS]{dispersion technology software}

\acro{ECD}[ECD]{equivalent circuit diagram}

\acro{FFL}[FFL]{field-free-line}
\acro{FFP}[FFP]{field-free-point}
\acro{FFR}[FFR]{field-free-region}
\acro{FFT}[FFT]{fast Fourier transform}
\acro{FOV}[FOV]{field-of-view}
\acro{FPGA}[FPGA]{Field Programmable Gate Array}
\acro{FWHM}[FWHM]{full width at half maximum}

\acro{GUI}[GUI]{graphical user interface}

\acro{HCC}[HCC]{high current circuit}

\acro{ICN}[ICN]{inductive coupling network}
\acro{ISI}[ISI]{integrated signal intensity}
\acro{ICs}[ICs]{integrated circuits}
\acro{IA}[IA]{instrumentation amplifier}
\acro{ICU}[ICU]{intensive care unit} 

\acro{LC}[LC]{inductor-capacitor}
\acro{LFR}[LFR]{low-field-region}
\acro{LNA}[LNA]{low noise amplifier}
\acro{LMA}[LMA]{Levenberg-Marquardt algorithm}

\acro{MPI}[MPI]{magnetic particle imaging}
\acro{MRI}[MRI]{magnetic resonance imaging}
\acro{MTT}[MTT]{mean-transit-time}
\acro{MNP}[MNP]{magnetic nanoparticle}
\acro{MPS}[MPS]{magnetic particle spectroscopy}
\acro{MAE}[MAE]{mean absolute error}
\acro{MSE}[MSE]{mean squared error}
\acro{MMR}[MMR]{magneto-mechanical resonator}
\acro{MEMS}[MEMS]{micro-electromechanical systems}
\acro{MMT}[MMT]{magneto-mechanical transmitter}

\acro{NLLS}[NLLS]{non-linear least squares}

\acro{PSF}[PSF]{point spread function}
\acro{PNS}[PNS]{peripheral nerve stimulation}
\acro{PTT}[PTT]{pulmonary transit time}
\acro{PDI}[PDI]{polydispersity index}

\acro{Q}[Q-factor]{quality factor}

\acro{RF}[RF]{radio-frequency}
\acro{RP}[RP]{RedPitaya STEMlab 125-14}
\acro{RF-fields}[RF]{radio frequency fields}
\acro{rBV}[rBV]{relative blood-volume}
\acro{rBF}[rBF]{relative blood-flow}
\acro{rCBV}[rCBV]{relative cerebral-blood-volume}
\acro{rCBF}[rCBF]{relative cerebral-blood-flow}
\acro{RBCs}[RBCs]{red blood cells}

\acro{SNR}[SNR]{signal-to-noise ratio}
\acro{SU}[SU]{surveillance unit}
\acro{SAR}[SAR]{specific absorption rate}
\acro{SPIONs}[SPIONs]{superparamagnetic iron-oxide nanoparticles}
\acro{USPIONs}[USPIONs]{ultrasmall superparamagnetic iron-oxide nanoparticles}
\acro{SF}{selection field}
\acro{SFG}[SFG]{selection-field generator}
\acro{SEM}[SEM]{scanning electron microscope}
\acro{SSIM}[SSIM]{structural similarity index measure}

\acro{TF}[TF]{transfer function}
\acro{THD}[THD]{total harmonic distortion}
\acro{TTP}[TTP]{time-to-peak}
\acro{TEM}[TEM]{transmission electron microscopy}
\acro{TxRx}[TxRx]{transmit-receive}

\acro{USPIO}[USPIO]{ultra-small superparamagnetic iron oxide}
\acro{ULF}[ULF]{ultra-low frequency}

\acro{VOI}[VOI]{volume of interest}
\acro{VSM}[VSM]{vibrating sample magnetometry}

\end{acronym}